# Emerging Two-dimensional Materials: graphene and its other structural analogues

*Gautam Mukhopadhyay*[*] *and Harihar Behera*

*Department of Physics, Indian Institute of Technology Bombay, Mumbai-400076, India*

*E-mail: gmukh@phy.iitb.ac.in

**Abstract.** The study of graphene, since its discovery around 2004, is possibly the largest and fastest growing field of research in material science, because of its exotic mechanical, thermal, electronic, optical and chemical properties. The studies of graphene have also led to further research in exploring the field of two dimensional (2D) systems in general. For instance, a number of other 2D crystals (not based on carbon, e.g., boronitrene, silicone, graphane, etc.) have been synthesized or predicted theoretically in recent years. Further, theoretical studies have predicted the possibility of other 2D hexagonal crystals of Ge, SiC, GeC, AlN, GaN, etc. The properties of these 2D materials are very different from their bulk. We shall present the general exotic properties of graphene like 2D systems followed by our computational results on the structural and electronic properties of some of them.

## 1. Introduction

Graphene [1-4] is a one-atom-thick crystal of $sp^2$-bonded C atoms arranged in a two-dimensional (2D) hexagonal lattice as in a single layer of graphite. Thousands/hundreds/few layers of graphene are detached from the graphite lead of a pencil (invented in 1656 in England [5]) when we use it to write on a paper, a fact that went unnoticed, although graphite has been known to the mankind for centuries. The unambiguous synthesis (by mechanical exfoliation of graphite), identification (by transmission electron microscopy (TEM)) and experimental determination of some of the exotic properties of graphene were reported first in 2004, by the Manchester group led by Andre Geim and Konstantin S. Novoselov [1], who earned Nobel Prize in 2010 for the "groundbreaking experiments regarding the two-dimensional material graphene". The detailed historical background and the status of graphene research as of December 2010, are given in the Nobel Lecture of Geim [6] and that of K.S. Novolesov [7], which trace the sporadic attempts of graphene-related research back to 1859 by the British chemist Benjamin Brodie [8]. As per Geim [6], Ulrich Hofmann and Hanns-Peter Boehm in 1962 [9] seem to have identified (by relative TEM contrasts) the thinnest possible fragments of reduced graphene oxide and some of them might be mono-layers of graphene (MLG). The term "graphene" was introduced by Boehm and his colleagues in 1986 [10], by joining the term "graph" derived from the word "graphite" with the suffix "ene" that refers to polycyclic aromatic hydrocarbons. Now graphene is regarded as the basic building block of graphitic materials (i.e. graphite = stacked graphene, fullerenes = wrapped graphene, nanotube = rolled graphene, graphene nanoribbon = nano-scale finite area sized rectangular graphene). These graphitic materials are classified as the allotropes of graphene (allotropes are different structural modifications of an element in the same phase of matter, e.g., different solid forms).

Because of its exotic mechanical, thermal, electronic, optical and chemical properties, such as the high carrier mobility, a weak dependence of mobility on carrier concentration and temperature, unusual quantum hall effect, hardness exceeding 100 times that of the strongest steel of same thickness and yet flexible (graphene can sustain strains more than 20% without breaking, graphene is brittle at certain strain limit), high thermal conductivity comparable to that of diamond and 10 times greater than that of copper, negative coefficient of thermal expansion over a wide range of temperature, has potentials for many novel applications [11-15].

The synthesis of graphene was surprising, because the existence of free-standing 2D crystals was believed impossible for several years, because they would ultimately turn into three-dimensional (3D) objects as predicted by Peierls [16], Landau [17] and Mermin [18]. The atomic displacements due to thermal fluctuations may be of the same order of magnitude as the inter-atomic distances, making the crystal unstable. Further, experiments indicate that thin films cannot be synthesized below a certain thickness, due to islands formations or even decomposition. Thus the synthesis of graphene [1] put a question mark on these predictions [16-18]. However, it was shown experimentally [19] that free-standing graphene sheets display spontaneous ripples owing to thermal fluctuations, and therefore real graphene is not perfectly flat.

The studies of graphene have also led to further research in exploring the field of two dimensional (2D) systems in general. Attempts have been made to understand the physics and chemistry of graphene on the basis of the *ab initio* calculations based on the density functional theory which leads to the so-called mass-less Dirac energy spectrum for electrons in graphene, although most of exotic properties are derivable from effective Hamiltonian for mass-less Dirac electrons, obtainable [12-14] from tight binding model. We will focus on our *ab initio* calculations of the structural and electronic properties of graphene in detail, then briefly state the effective Hamiltonian for mass-less Dirac electrons in graphene and discuss the linear energy dispersion as obtainable from this effective Hamiltonian as we proceed here. The studies of graphene have also led to further research in exploring the field of two dimensional (2D) systems in general. For instance, a number of other 2D crystals (not based on carbon) have been synthesized or predicted theoretically in recent years. Representative samples of other 2D nanocrystals which have been synthesized are boronitrene (graphene analogue of BN, the so called "white graphene", because of its color) [2, 20-26], silicene (silicon analog of graphene) [24-25, 27-32], graphane (fully hydrogenated graphene: extended 2D hydrocarbon) [33-34], composite structures of graphene and h-BN [35-37], ZnO [38-42], $MoS_2$ [2, 7, 27, 43], $NbSe_2$ [2,7,27], $Bi_2Sr_2CaCu_2O_x$ [2,7], etc. On the other hand, theoretical studies predict the possibility of other 2D hexagonal crystals of Ge [24, 25, 27, 32, 44-45], SiC [24, 25], GeC [24], AlN [24, 26], GaN [24], etc. The properties of some of these 2D materials whose electronic properties are significantly different from their bulk properties as revealed either by experiments or theory are listed in Table 1.

**Table 1.** Electronic properties of some 2D materials compared with their bulk properties

| Material | 2D | 3D |
| --- | --- | --- |
| C | Zero gap semiconductor | Semi-metal (graphite); Semicond. (Diamond) |
| Si | Zero gap semiconductor (PL-Si) | Semiconductor |
|  | Narrow gap semicond. (BL-Si) |  |
| Ge | Poor metal (PL-Ge) | Semiconductor |
|  | Zero gap semiconductor (BL-Ge) |  |
| SiC(hex) | Semicond., direct band gap | Semicond., indirect band gap |
| $MoS_2$ | Semicond., direct band gap | Semicond., indirect band gap |
| $NbSe_2$ | metal | CDW, superconductor |

With graphene as our model 2D system, we shall discuss the structural and electronic properties of some other graphene-like 2D systems based on our computational results.

## 2. Structural and electronic properties of graphene

One of the most surprising and interesting properties of graphene is its electronic band structure from which the so-called mass-less Dirac energy spectrum for electrons in graphene is derived. Elemental C has four valence electrons of which 2 electrons occupy the 2s orbital making it full filled, the $2p_x$ and the $2p_y$ orbitals containing 1electron each are half filled and the $2p_z$ orbital is empty. When C atoms come closer to form a crystal, one of the 2s electrons is excited to the $2p_z$ orbital from the energy gained from the neighboring nuclei; this has the net effect of lowering the overall energy of the system; interactions or bonding between the 2s and 2p orbitals of neighboring C atoms subsequently lead to the formation of new orbitals which are called hybrid orbitals; hybridized bound states of C atoms are then formed and due to the existence of multiple types of hybridization in C, different allotropes of C are formed as listed in Table 1. In graphene, the 2s and the two of the p orbitals ($2p_x$ and $2p_y$) interact covalently to form three $sp^2$ hybrid orbitals. The interactions of these three $sp^2$ orbitals lead to the formation of three σ-bonds which are the strongest type of covalent bond. These strong σ-bonds are responsible for the great strength and mechanical properties of graphene. The electrons forming the σ-bonds are localized in the plane containing the C atoms in graphene. The covalent bonds formed by the $2p_z$ electrons are called the π-bonds. The electrons in π-bonds are weakly bound to the nucleus hence relatively delocalized in a direction perpendicular to the plane containing the C atoms in graphene. These delocalized electrons are responsible for the electronic properties of graphene. Figure 1 depicts the ball-and-stick model of a typical iso-atomic 2D hexagonal crystal such as the graphene in two different configurations; one is planar (PL) and the other one is buckled (BL: no longer true 2D). Figure 2 depicts the view of a typical heterobilayer such as the graphene on a planar 2D hexagonal BN in some particular configuration.

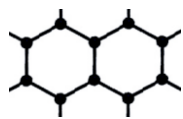 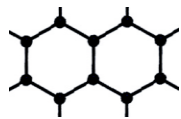

(a) Iso-atomic 2D-hex. Crystal Top view (PL)   (b) Iso-atomic 2D-hex. Crystal Top view (BL)

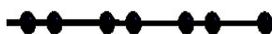 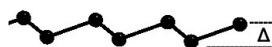

(c) Iso-atomic 2D-hex. Crystal side view (PL)   (d) Iso-atomic 2D-hex. Crystal Top view (BL)

**Figure 1**. Ball-and-stick model view of a typical iso-atomic 2D hexagonal crystal in two different structures. (a) Top view planar (PL), (b) Top view buckled (BL), (c) side view planar (PL) and (d) side view buckled. Buckling parameter Δ is the perpendicular distance between the two parallel planes in which the alternate atoms in a hexagon are positioned, for PL-structure Δ = 0.

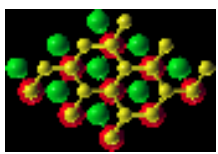 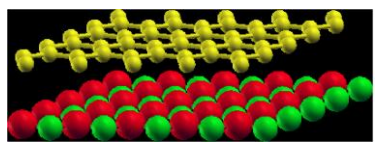

(a) A typical heterobilayer (Top view)   (b) A typical heterobilayer (Side view)

**Figure 2**. Ball-stick model view of a typical heterobilayer such as the graphene on 2D hex BN in certain configuration (B1): (a) Top view, and (b) Side view.

Graphene, has a 2D hexagonal lattice with two atoms per unit cell. A particular choice of the primitive lattice vectors may be made as: $a_1 = a(0.5 \times 3^{1/2}, 0.5)$, $a_2 = a(0.5 \times 3^{1/2}, -0.5)$ with $|a_1| = |a_2| = a$ (the lattice constant of graphene), the angle between $a_1$ and $a_2$ is 60°. In the lattice coordinates, the two C atoms are positioned at the points (0, 0, 0) and (2/3, 1/3, 0) in a planar structure and at the positions (0, 0, 0) and (2/3, 1/3, Δ), in a buckled structure. However, in the ab initio simulations of graphene, the 2D hexagonal structures are simulated using 3D super-cells with a large value of the "c" parameter ($a_3 = (0, 0, c)$), which ensures negligible interactions between the periodic images and the layers are effectively isolated. In this construction the atomic positions in the lattice coordinates are (0, 0, 0) and (2/3, 1/3, Δ/c) for a buckled structure, Δ = 0 for a planar structure. The unit cell of graphene has an area $|a_1 \bullet a_2| = a^2/2$. The carbon-carbon bond length $a_{C-C} = a/3^{1/2}$. Once the basis vectors of the real lattice are defined, the expression $a_i \bullet b_j = 2\pi \delta_{ij}$ ($\delta_{ij} = 1$ for i = j and $\delta_{ij} = 0$ for i ≠ j) allows us to calculate the reciprocal lattice vectors labeled as $b_j$ (j = 1, 2). The angle between $b_1$ and $b_2$ is 120°. The coordinates of the high symmetry points are Γ = (0, 0), M = $(2\pi/a)(1/3^{1/2}, 0)$ and K = $(2\pi/a)(1/3^{1/2}, 1/3)$. The electronic band structure of graphene can be described by a tight-binding model [56] which gives the energy of the electrons with wave vector **k** (= $k_x$, $k_y$) as

$$E = \pm\sqrt{\gamma_0^2 \left(1 + 4\cos^2\frac{k_y a}{2} + 4\cos\frac{k_y a}{2} \cdot \cos\frac{k_x\sqrt{3}a}{2}\right)},$$

where $\gamma_0$ is the nearest neighbor hopping energy integral, and $a$ is the lattice constant. The energy band dispersion is generally calculated along the perimeter of the triangle formed by these three points Γ, M and K (Figure -3).

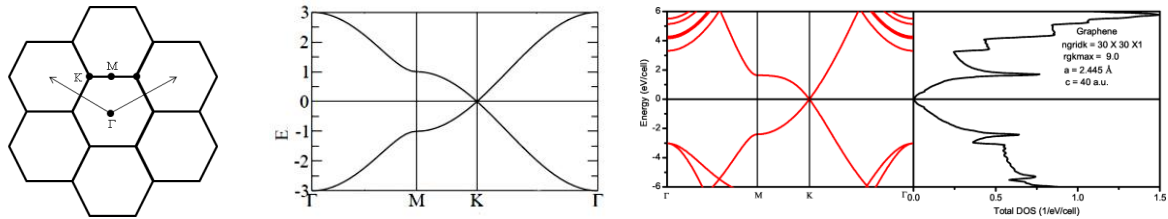

Figure 3. Left figure shows the high symmetry points Γ, M, and K of the hexagonal Brillouin zone. Note that there are six K point and six M points in the hexagonal BZ related by specific translational and rotational symmetry. The two arrows represent two reciprocal lattice vectors $b_1$ and $b_2$. The middle figure shows the tight binding energy bands (E in unit of $\gamma_0$). The figure in right shows our calculated band structures and total density of states (DOS) of graphene.

We use the density functional theory (DFT) based full-potential (linearized) augmented plane wave plus local orbital (FP+(L)APW+lo) method [46-47] as implemented in the elk-code [48] for our calculations. All our calculations discussed here are based on local density approximation (LDA) [49]. In Figure 4, we depict our investigation of the buckling in graphene (using the energy minimization procedure) and the bands and DOS in the vicinity of the K point of the BZ in its ground state (T = 0). As seen in Figure 4, minimum energy correspond*s* to $\Delta = 0$ and hence graphene has a planar structure in its ground state. This result is not in conflict with Landau-Mermin theory of the stability of 2D crystals as T = 0 here. The bands touch each other at the Fermi level ($E_F = 0$). The Hamiltonian that describes the electronic properties of graphene near the Fermi level can be approximated as [13, 55]: $H = v_F \sigma \bullet p = \hbar v_F(\sigma_x k_x + \sigma_y k_y)$, where $\sigma$ is a 2D pseudo-spin matrix ($\sigma_x$ and $\sigma_y$ are the x- and y-components of the Pauli spin matrices) describing the two sub lattices of the honeycomb lattice (pseudospin is an index that indicates on which of the two sub lattices a quasiparticles is located [13]), **p** is the momentum operator and $v_F$ (the Fermi velocity of the charge carriers) plays the role of "c" as in the Dirac Hamiltonian of a mass-less particle. If, $\phi_A$ and $\phi_B$ are the amplitudes of the wave function on sub lattices A and B, then the two component spinor $\psi$ is represented as a column matrix with $\phi_A$ and $\phi_B$ as its elements. Hamiltonian H acting on $\psi$ yields the Dirac-like linear dispersion relation $E_\pm = \pm \hbar k v_F$ (Figure 4). The positive energy and negative energy solutions, which correspond to conduction and valence bands respectively, meet at k = 0, implying the absence of band gap. Thus the electrons and holes in graphene mimic the mass-less Dirac fermions' behavior. The touching point is called the Dirac point $E_D$ in the literature. The charge carriers in graphene are called mass-less Dirac fermions and may therefore be described by a 2D analog of the Dirac equation for mass-less fermions. In the tight-binding (TB) study of graphene [12], one has $v_F = (a\gamma_0 3^{1/2})/2\hbar$. The precise value of $\gamma_0$ is difficult to determine analytically; as such, it is often used as a fitting parameter to match the ab initio computations or experimental data. Commonly used values for $\gamma_0$ range from about 2.7 to 3.3 eV. For routine calculations, the extracted value $\gamma_0 = 3.1$ eV from experimental measurements of $v_F \approx 10^6$ m/s in graphene is used [54]. In TB calculations, most authors use the in-plane lattice constant of graphite $a = 2.46$ Å as the "*a*" value for graphene. However, with our LDA value of $a = 2.445$ Å and $v_F = 0.83 \times 10^6$ m/s, our estimated value of $\gamma_0$ is 2.59 eV. Our calculated value of the $v_F$ is in close agreement with the experimentally observed value of $v_F = 0.79 \times 10^6$ m/s in graphene monolayer deposited on graphite substrate [50], (surprisingly) with the measured $v_F = 0.81 \times 10^6$ m/s in metallic carbon nano-tube [51] and about 20% less than that measured in coupled multi-layers [52-53] and the tight-binding value [12]. The reduction of $v_F$ in [50] was attributed to the electron-phonon interactions due to strong coupling with the graphite substrate.

However, the close agreement of our calculated result of free-standing graphene layer with results of [50] suggests a very weak coupling of graphene to the graphite substrate used in [50]. It is to be noted that, because of the weak coupling between the individual layers in graphite, extraction of graphene layer from graphite is possible by the scotch-tape method.

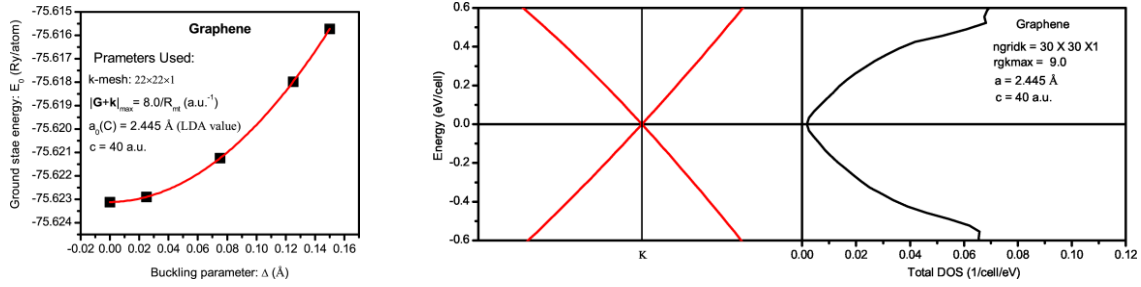

Figure 4. Left panel shows our probe of the buckling in graphene in ground state. Right panel shows the bands and DOS of graphene close to Fermi level (small value of DOS at Fermi level is due to our use of insufficient number of k points in the calculation of DOS).

## 3. Electronic properties of some other 2D crystals

In the following we will present and briefly discuss our calculated results on the ground state structural and electronic properties of some 2D hexagonal materials, viz., Silicene, Germanene, 2D h-BN and h-AlN, and Graphene/h-BN hetero-bilayer.

**Silicene:** Planar silicene (PL-Si) and Buckled silicene (BL-Si)

Table 2. Calculated structural and electronic properties of silicene compared with reported results.

| Silicene | $a_0$(Å) | $\Delta$(Å) | $v_F(10^5$ m/s) | $m^*$ ($m_e$) | $E_g$(meV) | Remark |
|---|---|---|---|---|---|---|
| PL-Si, (assumed) | 3.8453 | | 5.0 | 0.0 | 0.0 | This work (new calc.) |
| | 3.8454 | | | 0.0 | 0.0 | Our work [31] |
| | 3.86 | | | 0.0 | 0.0 | PAW-pot [27] |
| BL-Si | 3.8081 | 0.435 | 6.0 | 0.017 | 25.0 | This work (new calc.) |
| | 3.8081 | 0.44 | | | 0.0 | Our work [31], erroneous $E_g$ |
| | 3.8100 | 0.4247 | | | 0.0 | Pseudo-pot [25] |
| | 3.83 | 0.44 | | | 0.0 | PAW-pot [24] |

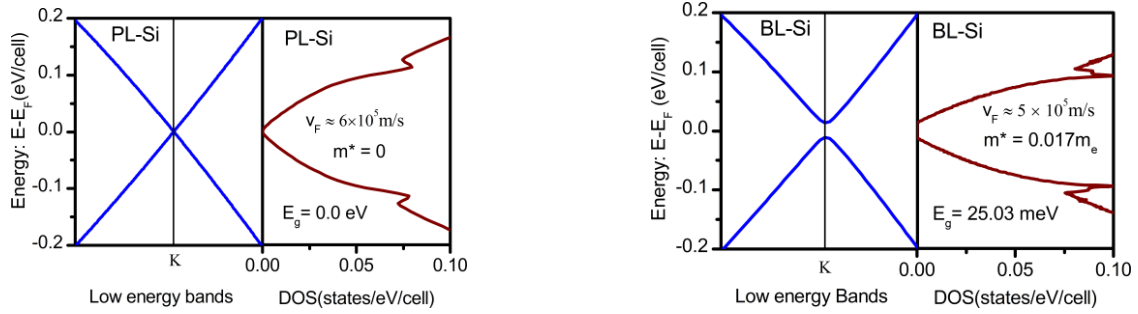

Figure 5. Low energy band dispersions at the K point and close to the Fermi level ($E_F = 0$) of PL-Si and BL-Si are compared along with their total DOS.

**Germanene**: Planar germanene (PL-Ge) and Buckled germanene (BL-Ge)

**Table3.** Calculated results of PL-Ge and BL-Ge compared with reported results.

| Germanene | $a_0$(Å) | $\Delta$(Å) | $v_F$($10^5$ m/s) | $E_g$(meV) | Remark |
|---|---|---|---|---|---|
| PL-Ge (assumed) | 3.999 | | 5.853 | Poor metal | This work |
| | 4.034 | | | Poor metal | PAW-pot [27] |
| BL-Ge | 3.9421 | 0.635 | 5.481 | 0.0 | Our work [44] |
| | 3.97 | 0.64 | | 0.0 | PAW-pot-[24] |
| | 3.9204 | 0.622 | | 0.0 | Pseudo-pot [25] |

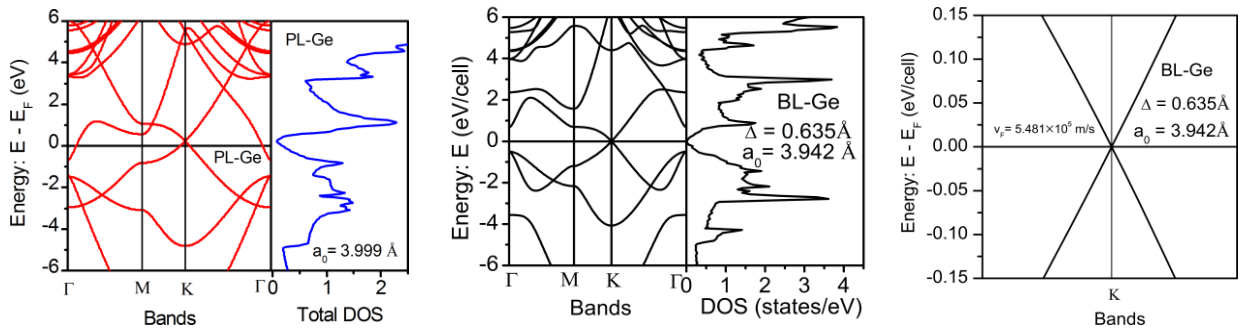

**Figure 6.** Bands and DOS of PL-Ge and BL-Ge compared; the rightmost panel shows the band dispersions of BL-Ge in a finer energy scale close to the Fermi level and at the K point of BZ.

## 2D h-BN and h-AlN

**Table 1**. Calculated structural parameters and energy band gaps of 2D h-BN and AlN listed with the reported values.

| Material | $a_0$(Å) | $\Delta$(Å) | $E_g$(meV) | Remark |
|---|---|---|---|---|
| 2D h-BN | 2.488 | 0.0 | 4.606 (KK) | Our work [26] |
|  | 2.488 | 0.0 | 4.613 (KK) | Li et. al. [36] |
|  | 2.51 |  | 4.61 (KK) | PAW-pot-[24] |
|  | 2.494 |  | 5.971 | Expt. [23] |
| 2D h-AlN | 3.09 | 0.0 | 3.037 (K$\Gamma$) | Our work [26] |
|  | 3.09 | 0.0 | 3.08 (K$\Gamma$) | PAW-pot-[24] |

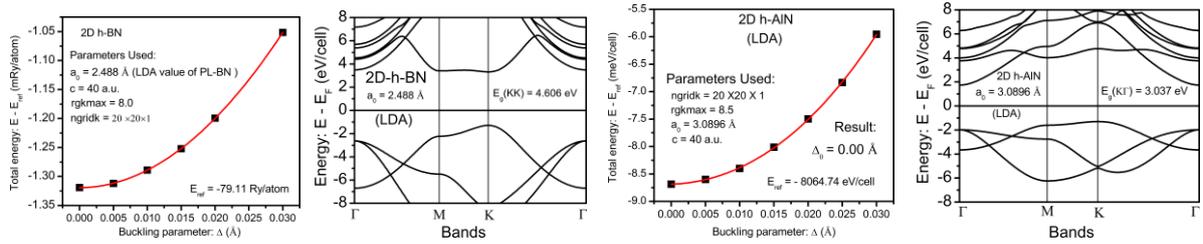

Figure 7. Determination of buckling parameter $\Delta$ and the band structure plots of 2D h-BN and AlN.

## Graphene/h-BN Hetero-bilayer

A system of graphene/h-BN heterobilayer can be regarded as the smallest and the thinnest form of BN gated graphene field effect-transistor (BN-GFET) that is feasible as per the recent experimental [35] and theoretical studies [36, 37]. As per theoretical calculations [36,37], graphene on h-BN monolayer or h-BN substrate forms stable a bound system in some particular configuration (B1) as shown in Figure 2. Recently, we have studied [37] such a bilayer system in detail and examined the effect of biaxial strain on this system. Here, we briefly note that graphene on a monolayer of h-BN or a h-BN substrate exhibits a band gap more than 50 meV (59 meV as per our LDA result) in a particular configuration (B1, see Figure 2) as per theoretical studies [36,37]. However, experimentally no such gap has been observed yet, and the absence of gap has been attributed to the misalignment of graphene layer with the BN substrate. The question of opening a band gap in graphene/h-BN system is now a technological challenge.

However, experimentally graphene/h-BN system shows superior electrical properties comparable to that of free standing graphene. There are several factors responsible for this such as the ultra-flat nature of h-BN substrate, reduced electron-phonon interactions, absence of charge traps as in oxide substrates. However, we have found another contributing factor (the Fermi velocity, $v_F$) for the freestanding graphene-like properties of graphene/h-BN system, from the band structure calculations of graphene and graphene/h-BN hetero-bilayer as shown in Figure 8.

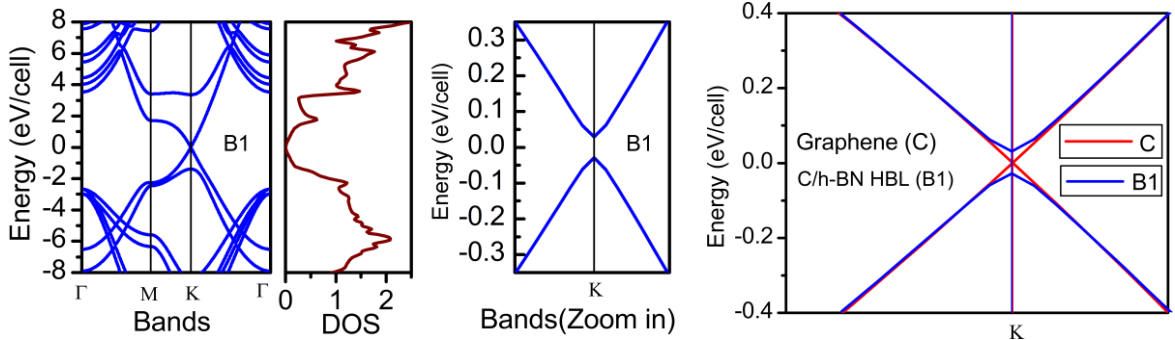

Figure 8. Band structures of graphene/h-BN heterobilayer (C/h-BN-HBL) showing a band gap; superimposed band of graphene and C/h-BN HBL (see the rightmost panel) shows both have same slopes and hence the same value of $v_F$.

## 4. Conclusions

We have briefly discussed an emerging trend in the study of 2D crystals which reveal interesting properties for potential applications. In particular, with graphene as the model system we have discussed the structural and electronic properties of some other 2D structural analogues of graphene such as silicene, germanene, h-BN, h-AlN and graphene/h-BN hetero-bilayer based on our current theoretical studies on these materials and compared our results with the reported ones. This provides us a flavor of the emerging new materials and their peculiar properties for potential novel applications. Understanding and tailoring the material properties of new and emerging materials at the nano-scale for desired applications are of great importance for fabrication and development of new and novel nano-devices. Considering that we are at the beginning of a new 2D era of material science, we should be interested for an entry into this new field of research, because the chances of success are high when a field is new.